\documentclass[aps,pra,twocolumn,floatfix]{revtex4-1}

\usepackage{amsfonts,amsmath,amssymb,amsthm,amscd}
\usepackage{acronym,array,bm,ccaption,color,dcolumn}
\usepackage{epsfig,graphicx,nomencl,revsymb4-1,upgreek,url}
\usepackage{hyperref}
\hypersetup{colorlinks=true, pdfauthor=Kevin C. Young, pdftitle=}

\newcommand{\tr}{{\rm Tr\thinspace}}

\newcommand{\ket}[1]{\ensuremath{\left\vert{#1}\right\rangle}}

\newcommand{\abs}[1]{\left\vert #1 \right\vert}

\renewcommand{\Re}{\operatorname{Re}}

\newcommand{\erf}[1]{Eq.~(\ref{#1})}

\newcommand{\haqc}{\ensuremath{H_{\rm AQC}(t)}}

\newcommand{\haqcb}{\bar H_{\rm AQC}(t)}
\newcommand{\hc}{\ensuremath{H_{\rm C}}}
\newcommand{\hb}{\ensuremath{H_{\rm B}}}
\newcommand{\uc}{\ensuremath{\mathcal{U}_{\rm C}}}
\newcommand{\ub}{\ensuremath{\mathcal{U}_{\rm B}}}

\newcommand{\hddt}{\tilde H_{\rm DD}(t)}

\newcommand{\prsec}[1]{\vspace{0.0cm}\noindent\emph{\textbf{#1}} --}

\begin{document}

\title{Unification and limitations of error suppression techniques\\ for adiabatic quantum computing}

\author{Kevin C.~Young}
\email[Electronic address: ]{kyoung@sandia.gov}
\affiliation{Department of Scalable \& Secure Systems Research (08961),
Sandia National Laboratories, Livermore, CA 94550}
\author{Mohan Sarovar}
\email[Electronic address: ]{mnsarov@sandia.gov}
\affiliation{Department of Scalable \& Secure Systems Research (08961),
Sandia National Laboratories, Livermore, CA 94550}

\date{\today}

\begin{abstract}
\notag \noindent
While adiabatic quantum computation (AQC) possesses some intrinsic robustness to noise, it is expected that a form of error control will be necessary for large scale computations. Error control ideas developed for circuit-model quantum computation do not transfer easily to the AQC model and to date there have been two main proposals to \textit{suppress} errors during an AQC implementation: energy gap protection and dynamical decoupling. Here we show that these two methods are fundamentally related and may be analyzed within the same formalism. We analyze the effectiveness of such error suppression techniques and identify critical constraints on the performance of error suppression in AQC, suggesting that error suppression by itself is insufficient for fault-tolerant, large-scale AQC and that a form of error \textit{correction} is needed. 

This manuscript has been superseded by the articles, ``\emph{Error suppression and error correction in adiabatic quantum computation I: techniques and challenges},''  arXiv:1307.5893, and ``\emph{Error suppression and error correction in adiabatic quantum computation II: non-equilibrium dynamics},''  arXiv:1307.5892.  
\end{abstract}

\maketitle

\noindent Adiabatic quantum computation (AQC) \cite{Farhi:2000tw,Farhi:2001ug} is often praised for its inherent robustness to both dephasing and energy relaxation, phenomena known to plague the majority of quantum computing paradigms.  Numerous studies, however, have shown that single qubit noise is capable of driving undesirable transitions from the adiabatic ground state; e.g. Refs. \cite{Chi.Far.etal-2001,Gai-2006,Tie.Sch-2007,Ami.Tru.etal-2009,Veg.Ban.etal-2010}. This observation prompted the formulation of three error suppression techniques, each  leveraging the error detecting properties of quantum stabilizer codes \cite{Got-1997}: energy gap protection (EGP)\cite{Jordan:2006jb}, in which the addition of the stabilizer generators to the system Hamiltonian causes errors to incur large energetic penalties; dynamical decoupling (DD)\cite{Lid-2008}, whereby stabilizer generators are applied periodically as unitary operators, refocusing errors much like traditional spin echos; and Zeno effect suppression \cite{PazSilva:2012kp}, which inhibits errors by frequent measurement of the stabilizer generators.  These three techniques apparently operate by very different physical mechanisms.  However, Facchi \textit{et al.} \cite{Fac.Lid.etal-2004} have shown that both Zeno suppression and DD may be viewed as limiting cases of a more general mathematical framework.  In this work, we show the DD and EGP may themselves be unified under a single formalism, and that these two methods are remarkably similar in their error suppression power. 

In conventional quantum computing models (e.g. the circuit-model) it is well understood that such error suppression techniques by themselves are insufficient for fault-tolerant quantum computing. From a thermodynamic perspective this is because error suppression alone does not provide a mechanism to remove the entropy generated by errors from the encoded system. Since the thermodynamic argument is independent of the computational model, it is expected that error suppression alone is insufficient for fault-tolerant quantum computing in the AQC model as well. We provide strong evidence to support this expectation using several arguments. 

An outline of the paper is as follows: first we introduce a formal correspondence between the EGP and DD error suppression techniques for AQC by considering their Hamiltonian implementation in an interaction picture. We then show using a simple argument that, unlike circuit model quantum computing, encoded AQC does not tolerate any excursions outside the codespace, and errors in general are unrecoverable. This places stringent requirements on error suppression and correction timescales in AQC. Finally, we derive a master equation of non-Lindblad form that describes encoded adiabatic evolution by exploiting properties of the stabilizer encoding. This master equation makes apparent the dynamical mechanism behind error suppression and also allows us to analyze the scaling limitations of error suppression. 

\prsec{Encoded AQC} 
\label{sec:intro}
Consider a Hamiltonian acting on a tensor product Hilbert space, $\mathcal{H}_\textrm{sys} \otimes \mathcal{H}_\textrm{env}$, describing a system undergoing adiabatic quantum evolution while coupled to an external environment/bath:
$	H(t) =  \haqc{} + \sum_{j=1}^{n_e} E_j\otimes B_j  + \hb.$
Here $H_{\rm AQC}$ acts on the system and performs the adiabatic evolution, $B_j$ and $E_j$ are bath operators and single qubit Pauli error operators, respectively, that define the system-environment interaction. $\hb$ is the bath Hamiltonian. Time-dependent eigenstates of $H_{\rm AQC}(t)$ may be labeled as $\ket{n,k}_t$ according to their principle quantum number $n$, index $k$ distinguishing any degeneracy, and the time $t$. 

We assume that the evolution is sufficiently slow that we may safely ignore diabatic transitions. All errors are then due to the system-bath interaction which is able to induce transitions from the adiabatically evolving ground state and therefore cause the computation to fail.  To suppress these transitions using any of the techniques mentioned above requires encoding the system in an error detecting stabilizer code \cite{Got-1997}. Typically, the code is chosen to be the smallest code for which each $E_j$ in the system-bath interaction anti-commutes with at least one of the stabilizer generators.  This encoding will enlarge the system's Hilbert space by a factor of $2^{N_g}$, where $N_g$ is the number of stabilizer generators of the code.  The physical operators, $\sigma_x, \sigma_y, \sigma_z$ in $H_{\rm AQC}$ are replaced by the code's logical operators, $\bar X, \bar Y, \bar Z$, and a time-dependent system control Hamiltonian composed of elements from the stabilizer group (specified later) is added to implement any desired error suppression.  The encoded Hamiltonian is then:
\begin{equation}
\label{eq:aqchambar}
	\bar H(t) =  \haqcb + \hc(t) + \sum_{j=1}^{N_e} E_j\otimes B_j + \hb 
\end{equation}
We have assumed that the system-bath interaction remains qualitatively the same after the encoding, but is extended to $N_e > n_e$ terms to correspond the larger system size.   
States of the encoded system may now be labeled by the same two quantum numbers as before, but with $N_g$ additional quantum numbers given by the eigenvalues of the stabilizer generators, $S_m\ket{n,k;s_1,s_2,\ldots,s_{N_g}}_t = s_m\ket{n,k;s_1,s_2,\ldots,s_{N_g}}_t$, where $S_m$ is a generator of the stabilizer group and $s_m = \pm 1$. We shall refer to the \emph{codespace} as the set of states for which all the stabilizer eigenvalues are $+1$.

For the following discussion it will be useful to define $\mathcal{U}_n(t_1,t_2) = \textrm{exp}_+(-i\int_{t_1}^{t_2} ds H_n(s))$, for $n \in \{ \textrm{AQC}, \textrm{B}, \textrm{C}\}$ ($+$ denotes positive time ordering although this is unnecessary except for $n=\textrm{AQC}$), and $\mathcal{U}_n(t) \equiv \mathcal{U}_n(t,0)$. Note that $[\bar{H}_\textrm{AQC}(s), \hc(s')]=0 ~~ \forall s,s'$, because $\bar{H}_\textrm{AQC}$ only contains logical operators of the code and $\hc$ only contains stabilizer terms, and hence these unitaries all commute with each other. The following notation is used for an operator $A$ in an interaction picture with respect to the control: $\tilde{A}(t) \equiv \mathcal{U}^\dagger_\textrm{C}(t)\mathcal{U}^\dagger_\textrm{B}(t) A~ \mathcal{U}_\textrm{C}(t)\mathcal{U}_\textrm{B}(t)$, which is typically called the \textit{toggling frame}. Evolution of states in this frame is generated by the toggling frame Hamiltonian: $\tilde{H}(t) \equiv\uc^\dagger(t)\ub^\dagger(t)\left(\bar H(t) -\hc -\hb \right)\ub(t)\uc(t)$.

\prsec{Dynamical decoupling}
\label{sub:dynamical_decoupling}
Dynamical decoupling controls are chosen to sequentially apply the generators of the stabilizer group as unitary operators.
The stabilizers are applied in a particular order, given by the vector $\mathbf{n}$, at times given by $K(t)\in \mathbb{Z}$, so that at time $t$ the most recent operator applied to the system was $S_{\mathbf{n}_{K(t)}}$.  The unitary operator defining the toggling frame may then be written as,
$\uc^{\rm DD}(t) = \prod_{j=0}^{K(t)} S_{\mathbf{n}_j}$.  As a product of stabilizer generators, the operator, $\uc^\textrm{DD}(t)$, is an element of the full stabilizer group and therefore commutes with $\haqcb$.  In the toggling frame the encoded Hamiltonian takes the form:
\begin{align}
	\hddt
		 \notag
		&= \bar H_{\rm AQC}(t) + \sum_{j=1}^{N_e} \tilde E_j^{\rm DD}(t) \otimes \tilde B_j(t)
\end{align}
And because $E_j$ must either commute or anti-commute with each member of the stabilizer group \cite{Got-1997}, we write:
\begin{equation}
\label{eq:ejdd}
	\tilde E_j^{\rm DD}(t) = \uc^{\rm DD \dagger}(t) E_j \uc^{\rm DD}(t) = (-1)^{p(t)} E_j 
\end{equation}
where $p(t) = 0$ if $[E_j,\uc^{\rm DD}(t)]=0$ and $p(t) = 1$ if $\{E_j,\uc^{\rm DD}(t)\}=0$. An effective DD cycle is one that causes $p(t)$ to rapidly alternate between $+1$ and $-1$, resulting in the system-environment coupling being modulated by a rapidly oscillating function of $t$. The state of the combined system-plus-environment at time $t$ is given by: $\ket{\tilde \Psi(t)} = \exp_+( -i\int_0^t ds \tilde{H}_\textrm{DD}(s)) \ket{\tilde \Psi_0}$, with $\ket{\tilde \Psi_0}$, $\ket{\tilde \Psi} \in \mathcal{H}_\textrm{sys} \otimes \mathcal{H}_\textrm{env}$,  and we set $\hbar=1$ throughout. Integrating over the modulation factor $(-1)^{p(s)}$ in the exponential has the effect of suppressing the system-environment coupling over timescales longer than the DD inter-pulse period.  This intuition can be made more explicit by deriving the average Hamiltonian using a Magnus expansion\cite{Oteo00} whereupon the oscillating terms may be seen to cause a reduction in the average system-bath coupling\cite{Viola99}.

\prsec{Energy gap protection}
\label{sub:energy_gap_protection}
The EGP approach proceeds by setting the control Hamiltonian equal to a sum of stabilizer generators, $\hc^{\textrm{EGP}}(t) = -\alpha \sum_{m=1}^{N_g} S_m$, with $\alpha >0$. States in the codespace are then eigenstates of $\hc$ with eigenvalue $-\alpha N_{g}$, but any state outside the codespace is subjected to an energy penalty of at least $\alpha$ which is expected to reduce transitions into these states.  Since $\hc$ is a function of only the stabilizer generators, $\uc^{\rm EGP}(t)$ again commutes with the code's logical operators which comprise $\haqcb$, so we can write the Hamiltonian in the toggling frame as:
\begin{align}
	\notag
	\tilde H_{\rm EGP}(t) 
		&=  \haqcb + \sum_{j=1}^{N_e}  \tilde E_j^{\rm EGP}(t)\otimes \tilde B_j(t)
\end{align}
Error operators in the EGP toggling frame can be shown to take the form:
\begin{align}
	\tilde E_j^{\rm{EGP}}(t)
	\notag
	&=  E_j e^{\left(2 i \alpha t \sum_{\{S_m,E_j\}=0} S_m \right)}\\
	&=  e^{\left(-2 i \alpha t \sum_{\{S_m,E_j\}=0} S_m \right)} E_j,
	\label{eq:toggling_EGP}
\end{align}
where the sums are taken over all stabilizer generators $S_m$ that anti-commute with the error operator $E_j$.  To obtain this expression we have exploited the following: (i) the stabilizer generators commute with each other, and (ii) each generator either commutes or anti-commutes with the noise operators: $S_m E_j = \pm E_j S_m$.  Let $w_j$ be  the number of generators that anticommute with $E_j$.  Then the action of this toggling frame Hamiltonian on any state, $ \tilde{\ket{\Psi}}= \sum_i \tilde{\ket{\psi^i_c}}\otimes \tilde{\ket{\phi^i}} \in \mathcal{H}_\textrm{sys} \otimes \mathcal{H}_\textrm{env}$ with $\tilde{\ket{\psi^i_c}}$ in the codespace is, 
\begin{equation}
	\tilde H_{\rm {EGP}}(t)\tilde{\ket{ \Psi}} = \left(\haqcb + \sum_{j=1}^{N_e} E_je^{2 i w_j \alpha t}\otimes \tilde B_j(t) \right)\tilde{\ket{ \Psi}} \nonumber
\end{equation}
Thus the coupling term $E_j\otimes \tilde B_j(t)$ is modulated by a factor of $e^{2 i w_j \alpha t}$. The evolution of states in the codespace is described by exponentiating the integral of this Hamiltonian and integrating over the modulation factors above has the effect of suppressing the coupling of the system to the environment. Similar to DD, the error term is modulated by an oscillating function when acting on states in the codespace in the interaction picture. In the case of EGP the oscillations are smooth and sinusoidal whereas the oscillations for (impulsive) DD are square waves in time.  One may then mimic a decoupling sequence by choosing $\alpha$ so that the EGP oscillations match the frequency of a DD sequence.  Numerical studies provide evidence that in such cases EGP and DD both yield nearly identical evolutions of states in the codespace. 

This similarity motivates generalizations of EGP where the weight terms, $\alpha$, are not constant in time or equal across the stabilizer generators.  Many decoupling sequences vary the time interval between the pulses; UDD\cite{Uhrig07}, for example, chooses the pulse arrival times as $t_n = T \cos(n \pi/2(N+1))$, where $N$ is the total number of pulses in time interval $[0,T]$.  To mimic this UDD sequence, where the modulation frequency is not constant in time, we choose a time dependent weight term, $\alpha(t) = NT/\sqrt{t(T-t)}$.  This approach was used, in a slightly different context, by the authors of Ref.~\cite{Jones:2012tr} to produce an effective UDD sequence using continuous controls.  More generally, allowing $\alpha$ to vary in time allows the strongest identification between the DD and EGP approaches, and a unified treatment of both as quantum control protocols. For instance, choosing $\alpha_j(t) = \sum_i \pi \delta(t-t_i^j)/2$ applies $S_j$ at time $t_i^j$ as a unitary operator (impulsive DD), but in the EPG formalism.  Furthermore, this approach naturally lends itself to the application of optimal control techniques to choose  $\alpha_j(t)$ to optimally mitigate the system-bath interaction.



\prsec{Error suppression in AQC}
\label{sub:error_supp_aqc}
We pause to emphasize a key difference between encoded circuit-model quantum computation and encoded AQC. Typical analyses of the former consider any state in the \textit{correctable} subspace as uncorrupted since these states can be decoded perfectly at the end of the computation. However, this is not true for AQC. 
To illustrate this, we consider a simple case.  Suppose the system is encoded and initialized in the ground state $\ket{0}$ of an initial Hamiltonian, and evolves unperturbed under the adiabatically changing Hamiltonian until, at time $\tau$, a correctable Pauli error $E_j$ occurs.  Then the system evolves unperturbed through the end of the AQC, at which point we measure the code stabilizers.  Because the Hamiltonian always commutes with the stabilizers, the error can be detected and identified by its syndrome and thus we can return the system to the code space by applying $E_j$.  Unfortunately, in the timespan between the error and its subsequent correction, things go horribly awry.

The overall evolution according to our simplified error model is:
\begin{equation}
\label{eq:evsimp}
	\ket{\psi}_T = E_j \mathcal{U}_\textrm{AQC}(\tau, T) E_j \mathcal{U}_\textrm{AQC}(0,\tau) \ket{0}_0,
\end{equation}
where the unitary evolution generated by the adiabatic Hamiltonian is given by a time-ordered exponential,
\begin{equation}
\label{eq:uaqc}
\mathcal{U}_\textrm{AQC}(\tau, T) = \exp_+\left(-i\int_{\tau}^T \bar H_{\rm AQC}(s)ds \right).
\end{equation}
(We neglect the as-yet-unspecified error suppressing control Hamiltonian, as its presence does not change the result.)
We assume that the adiabatic algorithm is well implemented, so $\mathcal{U}_\textrm{AQC}(0,\tau) \ket{0}_0 = \ket{0}_{\tau}$.  Now, the encoded AQC Hamiltonian, $\bar H_{\rm AQC}(s)$, is a weighted sum of the code's logical $X$,$Y$, and $Z$ operators.  Each is a Pauli operator, so it either commutes or anticommutes with the error operator $E_j$.  The encoded AQC Hamiltonian (at any normalized time $s$) splits into commuting and an anticommuting terms, 
\begin{equation}
\label{eq:decomp}
	\bar H_{\rm AQC}(s) = \bar H_{j}^+(s)+\bar H_{j}^-(s),
\end{equation}
where $[\bar H_{j}^+(s), E_j] = \{\bar H_{j}^-(s), E_j\} = 0$.  After some algebra, Eq.~\eqref{eq:evsimp} becomes
\begin{equation}
\label{eq:evsimp2}
		\ket{\psi}_T = \exp_+\left(-i\int_{\tau}^T \left( \bar H_{\boldsymbol{\nu}}^+(s)-\bar H_{\boldsymbol{\nu}}^-(s)\right)ds \right) \ket{0}_\tau
\end{equation}
Between the time when the error happens ($\tau$) and when it is corrected ($T$), the encoded system experiences a new, \emph{effective} Hamiltonian,	
	$\bar H^\prime_j = \bar H_{j}^+(s)-\bar H_{j}^-(s) = \bar H_{\rm AQC}(s) - 2\bar H_{j}^-(s)$.  
Since the state $\ket{0}_\tau$ is not generally an eigenstate of $\bar H^\prime_j(\tau)$, the system will undergo unintended evolution within the code space, moving it out of the ground state (a logical error).

So, although $E_j$ is a detectable single-body Pauli error, $\bar{E_j}$ can induce logical errors that are undetectable by the code. Therefore we cannot recover the state at time $T$ with an application of a detection and correction operation even though the original error at $\tau$ was detectable. Error correction (or decoding) at the end of the computation is unlikely to be effective; correctable errors will be transformed to logical errors by the Hamiltonian over the course of the computation. This is analogous to an error during the implementation of a non-transversal gate in the circuit model. This scrambling of errors by $\haqcb$ places constraints on pure error suppression techniques in AQC because it means that the suppression has to be strong enough to keep the state in the code space (and not just in the correctable space). This motivates active error during the course of AQC evolution, though it also imposes strict conditions on the rate at which error correction must be performed.  Specifically, one must correct on timescales fast compared to the rate at which $\haqcb$ moves an erred state out of the correctable space.  We can estimate the correction timescale, $\tau_c$, by the demanding it to be fast compared to the norm of the induced error AQC Hamiltonian, \emph{i.e.,} $\left(\max_{j,s} \abs{\abs{\bar H_{j}^-(s)}}\right)\tau_c\ll1$.


\prsec{Master equation for encoded AQC} 
\label{sec:master}
We have shown that error suppression in AQC using DD or EGP are closely related techniques as far as codespace population dynamics is concerned. Now we formalize this further by deriving a master equation describing effective encoded adiabatic evolution when the qubits interact with an environment. 

We begin with the full Hamiltonian in \erf{eq:aqchambar}, and assume that the system-environment coupling is weak compared to the other terms in this Hamiltonian, and that the encoding has been chosen such that each $E_j$ is a detectable error. In addition to the weak coupling approximation we also employ the first Markov approximation \cite{Bre.Pet-2002}. Crucially, we do not employ the second Markov approximation \cite{Bre.Pet-2002} in the derivation and therefore the reduced dynamics is able to capture the modification of system-environment coupling, and hence decoherence, by controls such as dynamical decoupling or energy penalty terms. This allows us to derive the following master equation in the toggling frame \cite{SuppInfo}:
\begin{align}
\frac{\textrm{d}\tilde{\rho}(t)}{\textrm{d}t} = &-i[\bar{H}_\textrm{AQC}(t), \tilde{\rho}(t)] \nonumber \\
	&- \sum_{j=1}^{N_e} \int_0^t d\tau \bigg[ \bigg(C_{j}(\tau) \tilde{E}_j(t) \tilde{\Xi}_j(t,\tau)\tilde{\rho}(t) \nonumber \\
	& ~~~~~~~~~~~~~ - C_{j}^*(\tau) \tilde{E}_j(t) \tilde{\rho}(t) \tilde{\Xi}_j(t,\tau) \bigg) + \rm{h.c.} \bigg] \nonumber
\end{align}
where $\tilde{\Xi}_j(t,\tau) \equiv \mathcal{U}_\textrm{AQC}(t,t-\tau)~\tilde{E}_j(t-\tau)~\mathcal{U}^\dagger_\textrm{AQC}(t,t-\tau)$, and  $C_{j}(\tau) \equiv \tr_\textrm{env}\{\mathcal{U}{^\dagger}_\textrm{B}(\tau,0) B_j \mathcal{U}_\textrm{B}(\tau,0) B_j \sigma_\textrm{eq}\}$ is the bath correlation function. We have assumed that the environments of all the qubits are uncorrelated and stationary. Correlated environments can be captured by the same formalism but for simplicity we will not do so here. 

Now consider the change in the codespace population as a result of these dynamics. Let $\mathbf{P} = \prod_{m=1}^{N_g} \frac{1}{2}(\mathbf{I} + S_m)$ be the projector onto the codespace, and $\mathbf{Q} = \mathbf{I}-\mathbf{P}$. Then the change in the codespace population is $\frac{\textrm{d}P_c}{\textrm{d}t} = \tr\{\mathbf{P}\frac{\textrm{d}\tilde{\rho}}{\textrm{d}t}\mathbf{P}\}$. We can show that the dynamical equation for this code space population is well approximated by \cite{SuppInfo}:
\begin{align}
	\label{eq:pc_mastereq}
\frac{\textrm{d}P_c(t)}{\textrm{d}t} \approx 
	& 2\sum_{j=1}^{N_e} \int_0^t d\tau 
		\Re 
		\bigg\{ 
			C_{j}(\tau) m_j(t,\tau) \times \\ \notag
	&\tr \bigg(
			E_j \mathbf{P} \Xi_j(t,\tau) \mathbf{Q}  \tilde{\rho}(t) \mathbf{Q} -
			 E_j \Xi_j(t,\tau) \mathbf{P}\tilde{\rho}(t)\mathbf{P} \bigg)\!
		\bigg\}
\end{align}
where $\Xi_j(t,\tau) \equiv e^{-i\tau \bar H_\textrm{AQC}(t)} E_j~e^{i\tau \bar H_\textrm{AQC}(t)}$, and $m_j(t,\tau)$ is a modulation function that results from the control, and captures its influence on dissipation and decoherence. For the two cases of EGP and DD, this modulation function takes the form:
\begin{align}
m^\textrm{EGP}_j(t,\tau) &= e^{2i \alpha \tau w_j} \nonumber \\
m^\textrm{DD}_j(t,\tau) &= (-1)^{p(t)-p(t-\tau)}
\end{align}
These modulation functions are analogous to the \textit{filter functions} derived for describing dynamical decoupling for pure dephasing dynamics \cite{deSousa09}.
In the absence of the adiabatic evolution (i.e. when $\bar H_\textrm{AQC}=0$) the traces in this equation simplify to $\tr\{E_j \Xi_j(t,\tau) \mathbf{P}\tilde{\rho}(t)\mathbf{P}\} = \tr\{\mathbf{P}\tilde{\rho}(t)\mathbf{P}\}$, and $\tr\{E_j \mathbf{P} \Xi_j(t,\tau) \mathbf{Q}  \tilde{\rho}(t) \mathbf{Q}\} = \tr\{\mathbf{Q_1}\tilde{\rho}(t)\mathbf{Q_1}\}$, where $\mathbf{Q_1}$ is a projector onto the subspace of $\mathbf{Q}$ that contains states one error away from the codespace. This simplification allows the derivation of a classical master/rate equation for the codespace population:
\begin{align}
\frac{\textrm{d}P_c(t)}{\textrm{d}t} \bigg|_{H_\textrm{AQC}=0} &= - \sum_j  r^-_j(t) P_c(t) + \sum_j r^+_j(t) P_{e1}(t) \nonumber
\end{align}
where $r^+_j(t) \equiv 2\textrm{Re} \big\{ \int_0^t d\tau C_j(\tau) m_j(t,\tau) \big\}$, $r^-_j(t) \equiv 2\textrm{Re} \big\{ \int_0^t d\tau C_j(\tau) m^*_j(t,\tau) \big\}$, and $P_{e1}(t)$ is the population in the one-error subspace at time $t$. 

The effect of the control Hamiltonian can be clearly seen in this master equation. The rates $r^-_j(t)$ quantify the leakage out of the codespace per unit time. In the absence of a control Hamiltonian these rates are simply proportional to $\textrm{Re}\{\int_0^t d\tau C_j(\tau)\}$, a property of the environmental fluctuations alone. However the control Hamiltonian, in the case of DD and EGP, has the effect of modulating the integrand with an oscillating function and hence decreasing its amplitude. To illustrate this consider a classical approximation of the environment (e.g. the Kubo-Anderson stochastic model) for which the correlation function is purely real, and fix it to be exponentially decaying \footnote{This classical model of the bath will not capture relaxation and temperature effects correctly, however we consider it here because of the simple form of the resulting correlation function. A more complete calculation with a quantum correlation function is done in the Supplementary Information \cite{SuppInfo}.}, $C(t) \propto e^{-\gamma t}$ where $\gamma$ is the inverse correlation time. Consider EGP, where the modulation function is sinusoidal, for which, 
\begin{equation}
r^\pm_j(t) \propto \frac{2 \left(\gamma - \gamma e^{-\gamma t } \cos[2 \alpha w_j t]+2 \alpha w_j e^{-\gamma t }  \sin[2 \alpha w_j t]\right)}{(2 \alpha w_j)^2+\gamma ^2} \nonumber.
\end{equation}
In this expression, the consequences of adding the energy penalty terms are summarized by the presence of the factor $2\alpha w_j$, which increases with the energy penalty ($\alpha$) and the number of that anti-commute with the error ($w_j$). The term has two effects: (i) its presence in the denominator decreases the overall rate of population leakage, (ii) it increases the oscillation frequency of the sinusoidal functions in the numerator, thus decreasing the magnitude of integrals of $r^\pm_j(t)$. Therefore, this calculation explicitly shows how the control Hamiltonian decreases population leakage from the codespace. Note that it is possible to achieve a suppression of population leakage from the codespace that is exponential in the number of qubits by adding the whole stabilizer group as the penalty Hamiltonian -- i.e. in this case exactly half of the stabilizer group \cite{SuppInfo} will anti-commute with each error and $w_j = 2^{N-n_l-1} ~~\forall j$, where $N$ is the number of physical qubits and $n_l$ is the number of logical qubts. Alternatively one could scale the energy penalty to be exponential in the number of qubits: $\log \alpha \sim \mathcal{O}(N)$. However, both approaches require exponential resources because exponential energy must be added to the Hamiltonian (either via the penalty $\alpha$ or via the number of stabilizer terms added). It is an interesting question as to whether such exponential suppression is possible without this concomitant cost. For this model of noise (weak and not strongly non-Markovian) we conjecture that this is not possible.

Description of the dynamics as a classical rate equation is only possible when $\bar H_\textrm{AQC}=0$. Although a rate equation for the populations is not possible in this case because each state in the codespace has a different rate of leakage to the errors spaces, it is clear that the mechanism by which the control suppresses leakage from the codespace remains the same. We would ideally like $\textrm{d}P_c(t)/\textrm{d}t=0$ and since the trace quantities will not be zero in general, the alternative is to suppress the magnitude of the integral coefficients. This is exactly what the control Hamiltonian does by adding oscillatory components to the integrand defining the population transition rates, although the expressions for the suppressed leakage rates are not as simple as in the $H_\textrm{AQC}=0$ case.

\prsec{Discussion}
A practical consequence of the unification of the DD and EGP error suppression techniques is that it implies that DD can be used to emulate EGP. This is significant since EGP typically requires the addition of many-body terms (terms with a non-identity Pauli on more than two qubits of the system) to the system Hamiltonian, which is usually impractical. In addition, as illustrated above this unification further prompts generalizations of EGP where the penalty Hamiltonian coefficients ($\alpha$) are not uniform and time dependent. Finally, our analysis of the limitations of error suppression in AQC motivates schemes for error correction within this model of quantum computation, while we have also shown that implementations of error correction are non-trivial due to the always-on adiabatic evolution. In a follow-up paper, motivated by these observations, we will formulate methods of error correction that are compatible with AQC.

Note that this manuscript has been superseded by that of Refs.~\cite{Young2013,Sarovar2013}.

\prsec{Acknowledgements} 
\label{sec:acknowledgements}
We acknowledge important discussions with Sandia's AQUARIUS Architecture team, especially with Robin Blume-Kohout and Anand Ganti. This work was supported by the Laboratory Directed Research and Development program at Sandia National Laboratories. Sandia is a multi-program laboratory managed and operated  by Sandia Corporation, a wholly owned subsidiary of Lockheed Martin Corporation, for the United States Department of Energy's National Nuclear Security Administration under contract DE-AC04-94AL85000.

\bibliography{aqc_error_suppression}

\begin{widetext}
\numberwithin{equation}{subsection}
\pagebreak
\section*{Supplementary Information}
\subsection{Appendix: Detectable errors commute with exactly half of the stabilizer group}
\noindent\emph{Theorem}: All detectable error operators of an error detecting code must anti-commute with exactly half of the full stabilizer group and commute with the other half. \\

\noindent\emph{Proof}: The stabilizer group, $\mathcal G$, of an error detecting code is an abelian group generated by a set of stabilizer generators, $\mathcal{S}$.  The set of detectable errors, $\mathcal{E}$, is composed of all operators which anti-commute with at least one of the stabilizer generators. Because both the stabilizer group and the set of correctable errors are contained within the $N$-qubit Pauli group, all such operators either commute or anti-commute with one another.  

Any given detectable error operator, $E_j$, will partition $\mathcal G$ into commuting and anti-commuting subsets $\mathcal{G}_j^{\pm} \subset \mathcal{G} = \{g \in \mathcal{G} \vert E_j g = \pm g E_j\}$.  Because $E_j$ is detectable, we can choose a generator, $S_0^j$, with which it anti-commutes.  This generator defines an involutive map between the sets $\mathcal{G}_j^\pm$ by $g\in \mathcal{G}_j^\pm \rightarrow S_0^j g\in \mathcal{G}_j^\mp$.  That is, if $g$ commutes with $E_j$, then $S_j^0 g$ anticommutes with $E_j$, and visa-versa.  The sets $\mathcal{G}_j^\pm$ are then isomorphic, so  are the same size.  

\subsection{Appendix: Derivation of encoded AQC master equation}
\label{sec:me_app}

In this Appendix we outline the derivation of a master equation describing the effective adiabatic evolution when the qubits are coupled to uncontrolled (environmental) degrees of freedom. By employing fewer approximations than in the derivation of the conventional Lindblad master equation this reduced dynamics is able to capture the modification of system-environment coupling, and hence decoherence, by controls such as dynamical decoupling or energy penalty terms.

We begin with a Hamiltonian acting on the combined Hilbert space of encoded system and environment ($\mathcal{H}_\textrm{sys} \otimes \mathcal{H}_\textrm{env}$) of the form:
\begin{equation}
H(t) = \bar{H}_\textrm{AQC}(t) + \hc(t) +  \sum_{j=1}^{N_e} E_j\otimes B_j + H_B
\end{equation}
where all these quantities are defined in the main text (\erf{eq:aqchambar}).
Define an interaction picture with respect to $\bar{H}_\textrm{AQC}(t)$, $H_C(t)$ and $H_B$ as: $\breve{A}(t) \equiv \mathcal{U}^\dagger(t,0) ~A~ \mathcal{U}(t,0)$, where
\begin{equation}
\mathcal{U}(t,0) = e_+^{-i \int_0^t ds \bar{H}_\textrm{AQC}(s) + H_C(s) + H_B}
\end{equation}
and the subscript $+$ indicated positive time ordering of the expoential. A particularily important property of these Hamiltonians, which we will utilize later, is that they all commute. That is, $[\bar{H}_\textrm{AQC}(s), H_C(s')]=0 ~~~ \forall s,s'$, because $\bar{H}_\textrm{AQC}$ only contains logical operators and $H_C$ only contains stabilizer terms. And obviously $H_B$ commutes with the other two terms. This property implies that this interaction picture transformation factors into: $\mathcal{U}(t,0) = \prod_n \mathcal{U}_n(t,0) = \prod_n e_+^{-i \int_0^t ds H_n(s)}$ with $n \in \{\textrm{AQC}, C, B\}$. 

Let $\varrho$ be the combined density matrix of system and environment, i.e. is a normalized trace-class operator in $\mathcal{H}_\textrm{sys}\otimes \mathcal{H}_\textrm{env}$. A formal solution to the von-Neumann equation in the interaction picture, $\frac{\textrm{d}\breve{\varrho}(t)}{\textrm{d}t} = -i[\breve{H}(t), \breve{\varrho}(t)]$, is \cite{Bre.Pet-2002}:
\begin{equation}
\frac{\textrm{d}\breve{\varrho}(t)}{\textrm{d}t} = -i[ \breve{H}_I(t), \breve{\varrho}(0)] - \int_0^t [\breve{H}_I(t), [\breve{H}_I(s), \breve{\varrho}(s)]] ds
\end{equation}
where $H_I \equiv \sum_{j=1}^{N_e} E_j\otimes B_j$.
We will assume that the weak system-environment coupling does not perturb the environment from its equilibrium state at timescales that we resolve, and hence $\breve{\varrho}(s) \approx \breve{\rho}(s) \otimes \sigma_\textrm{eq}$, a tensor product of the system density matrix $\breve{\rho}(s) \equiv \tr_\textrm{env} \{\breve{\varrho}(s)\}$, and the environmental equilibrium density matrix. This allows us to derive a time-convolution master equation for the system density matrix \cite{Bre.Pet-2002}:
\begin{align}
\frac{\textrm{d}\breve{\rho}(t)}{\textrm{d}t} = &\tr_\textrm{env}\{\frac{\textrm{d}\breve{\varrho}(t)}{\textrm{d}t} \} \nonumber \\
	= &-\sum_{j,k=1}^{N_e} \int_0^t ds \bigg(C_{kj}(t,s) \breve{E}_k(t)\breve{E}_j(s)\breve{\rho}(s)
	      - C_{jk}(s,t) \breve{E}_k(t) \breve{\rho}(s) \breve{E}_j(s) \nonumber \\
	&  ~~~~~~~~~~~~~~~~~  - C_{kj}(t,s) \breve{E}_j(s) \breve{\rho}(s) \breve{E}_k(t)
	      + C_{jk}(s,t) \breve{\rho}(s) \breve{E}_j(s)\breve{E}_k(t) \bigg)
\end{align}
with $C_{kj}(t,s) \equiv \tr_\textrm{env}\{\breve{B}_k(t)\breve{B}_j(s)\sigma_\textrm{eq}\}$ is the quantum correlation function of the environment. To obtain this expression we have assumed that $\tr_\textrm{env}\{ B_j(t) \sigma_\textrm{eq}\} = 0 ~~\forall j$ -- i.e. the average interaction force on the bath equilibrium state is zero. We assume that the environment is stationary, implying that this correlation function is only dependent on the time difference $\tau = t-s$. This simplifies the master equation to:
\begin{align}
\frac{\textrm{d}\breve{\rho}(t)}{\textrm{d}t} = &-\sum_{j,k=1}^{N_e} \int_0^t d\tau \bigg(
		C_{kj}(\tau) \breve{E}_k(t)\breve{E}_j(t-\tau)\breve{\rho}(t-\tau) 
		- C_{kj}^*(\tau) \breve{E}_k(t) \breve{\rho}(t-\tau) \breve{E}_j(t-\tau) \nonumber \\
		&~~~~~~~~~~~~~~~~~~~~ - C_{kj}(\tau) \breve{E}_j(t-\tau) \breve{\rho}(t-\tau) \breve{E}_k(t)  
		+ C_{kj}^*(\tau) \breve{\rho}(t-\tau) \breve{E}_j(t-\tau)\breve{E}_k(t) \bigg)
\end{align}
The final approximation we make is typically referred to as the first Markov approximation and replaces $\breve{\rho}(t-\tau)$ with $\breve{\rho}(t)$ in the integrals above \cite{Bre.Pet-2002}. This amounts to assuming that the change in the system state (in the interaction picture, and therefore due to the weak system-environment coupling) is negligible on the timescale set by the decay of the environment correlation function. Therefore this formalism is valid for moderately fast-relaxing or weakly-coupled environments. Finally, we will restrict out analysis to uncorrelated environments for the system qubits, that is $C_{kj}(\tau) = \delta_{kj}C_j(\tau)$. The analysis that follows can be generalized to correlated environments but we will not do so here. 

We rewrite this resulting master equation in an interaction picture with respect to the control Hamiltonian only. In the dynamical decoupling literature this is known as the toggling frame, and is defined by: $\tilde{A}(t) = \mathcal{U}^\dagger_C(t,0) ~A~\mathcal{U}_C(t,0)$. The transformation required to move into this frame is particularly easy in this case because as noted above the stabilizer properties result in the factoring of the full interaction picture transformation unitary. In the toggling frame:
\begin{align}
\frac{\textrm{d}\tilde{\rho}(t)}{\textrm{d}t} &= -i[\bar{H}_\textrm{AQC}(t), \tilde{\rho}(t)] \nonumber \\
	- &\sum_{j=1}^{N_e} \int_0^t d\tau \bigg( C_j(\tau) \tilde{E}_j(t) \tilde{\Xi}_j(t,\tau)\tilde{\rho}(t) 
	 - C_j^*(\tau) \tilde{E}_j(t) \tilde{\rho}(t) \tilde{\Xi}_j(t,\tau)
	 - C_j(\tau) \tilde{\Xi}_j(t,\tau) \tilde{\rho}(t) \tilde{E}_j(t)
	 + C_j^*(\tau) \tilde{\rho}(t) \tilde{\Xi}_j(t,\tau)\breve{E}_j(t) \bigg) \nonumber
\end{align}
where $\tilde{\Xi}_j(t,\tau) \equiv \mathcal{U}_\textrm{AQC}(t,t-\tau)~\tilde{E}_j(t-\tau)~\mathcal{U}^\dagger_\textrm{AQC}(t,t-\tau)$. 

Now consider the change in the codespace (the no-error subspace) population as a result of these dynamics. As in the main text we define $\mathbf{P}$ as the projector onto the codespace, and $\mathbf{Q} = \mathbf{I}-\mathbf{P}$. Then the change in the codespace population is $\frac{\textrm{d}P_c}{\textrm{d}t} = \tr\{\mathbf{P}\frac{\textrm{d}\tilde{\rho}}{\textrm{d}t}\mathbf{P}\}$. To evaluate this quantity, we will first insert identities in the form $\mathbf{P+Q}$ around $\tilde{\rho}(t)$, resulting in:
\begin{align}
\frac{\textrm{d}P_c(t)}{\textrm{d}t} = - \tr \bigg\{ \sum_j \int_0^t d\tau 
		&C_j(\tau) \bigg[ \mathbf{P}\tilde{E}_j(t) \tilde{\Xi}_j(t,\tau) \mathbf{P}\tilde{\rho}(t)\mathbf{P} + \mathbf{P}\tilde{E}_j(t) \tilde{\Xi}_j(t,\tau) \mathbf{Q}\tilde{\rho}(t)\mathbf{P} \bigg] \nonumber \\
	-&C_j^*(\tau) \bigg[ \mathbf{P} \tilde{E}_j(t) \mathbf{Q} \tilde{\rho}(t)  \mathbf{Q} \tilde{\Xi}_j(t,\tau) \mathbf{P}\bigg] - C_j(\tau) \bigg[ \mathbf{P} \tilde{\Xi}_j(t,\tau) \mathbf{Q}  \tilde{\rho}(t) \mathbf{Q}  \tilde{E}_j(t) \mathbf{P} \bigg] \nonumber \\
	+&C_j^*(\tau) \bigg[\mathbf{P} \tilde{\rho}(t) \mathbf{P} \tilde{\Xi}_j(t,\tau)\tilde{E}_j(t) \mathbf{P} + \mathbf{P} \tilde{\rho}(t) \mathbf{Q} \tilde{\Xi}_j(t,\tau)\tilde{E}_j(t) \mathbf{P} \bigg] \bigg\}
\end{align}
where we have used the identities: $\mathbf{P}\mathbf{Q} = 0$, $\mathbf{P} \tilde{E}_j \mathbf{P} = 0 ~~\forall j$, and $\mathbf{P} \tilde{\Xi}_j(t, \tau) \mathbf{P} = 0 ~~ \forall j$. The first of these is by definition and the others follow from the properties of the Hamiltonian and error operators -- i.e. $\bar{H}_\textrm{AQC}(s)$ and $H_C(s)$ cannot move states between the subspaces defined by $\mathbf{P}$ and $\mathbf{Q}$, and $E_j$ applied to any state in $\mathbf{P}$ results in a state in $\mathbf{Q}$. The term $\tr\{ \mathbf{P}\tilde{E}_j(t) \tilde{\Xi}_j(t,\tau)\mathbf{Q}\tilde{\rho}(t)\mathbf{P}\}$ and its conjugate also evaluate to zero although it is slightly more involved to see why. The reason is that $\tilde{E}_j(t) \tilde{\Xi}_j(t,\tau) = \mathcal{U}^\dagger_C(t,0)E_j\mathcal{U}_C(t,0)\mathcal{U}_\textrm{AQC}(t,t-\tau) \mathcal{U}^\dagger_C(t-\tau,0)E_j\mathcal{U}_C(t-\tau,0)~\mathcal{U}^\dagger_\textrm{AQC}(t,t-\tau)$ contains two applications of $E_j$ interleaved with unitary evolution that does not connect different stabilizer sectors and hence cannot connect $\mathbf{P}$ and $\mathbf{Q}$ subspaces. Hence, this master equation simplifies to:
\begin{align}
\frac{\textrm{d}P_c(t)}{\textrm{d}t} = - \tr \bigg\{ \sum_j \int_0^t d\tau 
		&C_j(\tau) \bigg[ \mathbf{P}\tilde{E}_j(t) \tilde{\Xi}_j(t,\tau) \mathbf{P}\tilde{\rho}(t)\mathbf{P} \bigg]  + C_j^*(\tau) \bigg[\mathbf{P} \tilde{\rho}(t) \mathbf{P} \tilde{\Xi}_j(t,\tau)\tilde{E}_j(t) \mathbf{P} \bigg] \nonumber \\
	-&C_j^*(\tau) \bigg[ \mathbf{P} \tilde{E}_j(t) \mathbf{Q} \tilde{\rho}(t)  \mathbf{Q} \tilde{\Xi}_j(t,\tau) \mathbf{P}\bigg] - C_j(\tau) \bigg[ \mathbf{P} \tilde{\Xi}_j(t,\tau) \mathbf{Q}  \tilde{\rho}(t) \mathbf{Q}  \tilde{E}_j(t) \mathbf{P} \bigg] \bigg\}
\end{align}

At this point we employ a critical property of the control Hamiltonian: that it modulates the system-environment interaction. Using the expressions for toggling frame error operators in \erf{eq:ejdd} and \erf{eq:toggling_EGP} allows us to simplify the equation of motion for codespace population to:
\begin{align}
\frac{\textrm{d}P_c(t)}{\textrm{d}t} =  2\sum_j \int_0^t d\tau \Re \bigg\{ &C_j(\tau) m_j(t,\tau) \tr \bigg[ \mathbf{P} \hat{\Xi}_j(t,\tau) \mathbf{Q}  \tilde{\rho}(t) \mathbf{Q} E_j \mathbf{P} \bigg] \bigg\} - \Re \bigg\{  
	 C_j(\tau) m_j(t,\tau) \tr \bigg[ \mathbf{P}E_j \hat{\Xi}_j(t,\tau) \mathbf{P}\tilde{\rho}(t)\mathbf{P} \bigg]  \bigg\}  
\end{align}
where $\hat{\Xi}(t,\tau) \equiv \mathcal{U}_\textrm{AQC}(t,t-\tau)E_j~\mathcal{U}^\dagger_\textrm{AQC}(t,t-\tau)$. In this equation the effects of control (error suppression) are encapsulated in the modulation functions $m_j(t,\tau)$. For the two error suppression techniques, as shown in the main text, these functions take the form:
\begin{align}
m^\textrm{EGP}_j(t,\tau) &= e^{2i \alpha \tau w_j} \nonumber \\
m^\textrm{DD}_j(t,\tau) &= (-1)^{p(t)-p(t-\tau)}
\label{eq:mod_functions}
\end{align}
where $w_j$ is the number of stabilizer terms in the EGP penalty Hamiltonian that anti-commute with the error $E_j$. $p(t)$ is the DD coefficient defined in the main text. To write the modulation function for EGP we have exploited the property 
\begin{align}
e^{\left(2 i \alpha \tau \sum_{\{S_m,E_j\}=0} S_m \right)} \mathbf{P} = e^{2i \alpha \tau w_j } \mathbf{P}
\end{align}
which follows from the fact that all states in the codespace are eigenvalue $+1$ eigenstates of the stabilizers.

Since the correlation function decays with $\tau$, the value of the integrand at small values of $\tau$ are the most important. And if we assume that $\bar H_\textrm{AQC}$ varies slowly with respect to time, we can approximate
\begin{align}
\mathcal{U}_\textrm{AQC}(t,t-\tau) = e_+^{-i\int_{t-\tau}^t ds \bar H_\textrm{AQC}(s)} \approx e^{-i \tau \bar H_\textrm{AQC}(t)} 
\end{align}
and hence approximate $\hat{\Xi}(t,\tau) \approx \Xi(t,\tau) \equiv e^{-i\tau \bar H_\textrm{AQC}(t)} E_j~e^{i\tau \bar H_\textrm{AQC}(t)}$. Using this results in the final form of the population master equation used in the main text.

The modulation functions given in \erf{eq:mod_functions} display some degree of asymmetry between the EGP and DD error suppression techniques because while $m^\textrm{DD}$ depends on times $t$ and $\tau$, $m^\textrm{EGP}$ only depends on time $\tau$. We stress that this is only because we have restricted ourselves to the case of constant, uniform energy penalty $\alpha$. As mentioned in the main text, a more general formulation of EGP (from which a unification with DD is even more straightforward) would allow for $\alpha$ to be time dependent: $\hc^{\textrm{EGP}}(t) = - \sum_{m=1}^{N_g} \alpha_j(t) S_m$. In this case,
\begin{equation}
m^\textrm{EGP}_j(t,\tau) = (e^{2i})^{\chi(t)-\chi(t-\tau)}
\end{equation} 
with $\chi(t) \equiv \sum_{\{S_m,E_j\}=0} \int_0^{t} ds ~ \alpha_m(s)$. In this more general formulation the similarity between DD and EGP is even more evident.

\subsection{Appendix: Full rate calculation for Ohmic spectral density}
The example environment considered in the main text is a classical bath with exponentially decaying correlation. Here we generalize this to a true quantum environment and explicitly demonstrate that the controlled suppression of population leakage from the codespace holds in this case too. Consider a damped harmonic environment with an Ohmic spectral density with Lorentz-Drude regularization: $J(\omega) =  2 E_R \gamma \omega/(\omega^2 + \gamma^2)$, where $E_R$ is the ``reorganization energy" which quantifies the total system-environment coupling strength, and $\gamma$ is the inverse of the environment correlation timescale. The quantum correlation function for an environment with such a spectral density is:
\begin{equation}
C(t) = i2E_R \gamma \left( \frac{1}{e^{i\beta\gamma}-1}\right)e^{-\gamma t} - \sum_{\kappa=1}^\infty \frac{4 E_R \gamma}{\beta}\frac{\nu_\kappa}{\gamma_\kappa^2-\nu_\kappa^2}e^{-\nu_\kappa t} \nonumber
\end{equation}
where $\beta=1/k_BT$ is the inverse temperature and $\nu_\kappa \equiv \frac{2\pi \kappa}{\beta}$ are the \textit{Matsubara} frequencies. For fast decaying correlations (large $\gamma$) the terms in the summand decay quickly and it is customary to truncate the sum at finite $\kappa$. Assuming that the error suppression technique is EGP and computing the rate of leakage from the code space in the population master/rate equation yields:
\begin{align}
r^-_j(t) &= \frac{a_0 [ b_0 -\gamma e^{-\gamma t}\cos(-2\alpha w_j t - \frac{\beta\gamma}{2}) - 2\alpha w_j e^{-\gamma t}\sin(-2\alpha w_j t - \frac{\beta \gamma}{2}) ] }{(2\alpha w_j)^2 + \gamma^2} \nonumber \\
&~~~~ - \sum_{\kappa=1}^\infty \frac{a_\kappa [ b_\kappa - \nu_\kappa e^{-\nu_\kappa t}\cos(2\alpha w_j t ) + 2\alpha w_j e^{-\nu_\kappa t}\sin(2\alpha w_j t) ] } {(2\alpha w_j)^2 + \nu_\kappa^2}
\end{align}
and similarly for $r^-_j(t)$. Here $a_i, b_i$ are irrelevant constants. As in the case of a classical model of the environment, the suppression of these transition rates is achieved by two mechanisms: (i) the suppression term $2\alpha w_j$ in the denominator decreases the overall rate of population leakage, (ii) the same term increases the oscillation frequency of the sinusoidal functions in the numerator, thus decreasing the magnitude of integrals of $r^\pm_j(t)$. 

\end{widetext}



\end{document}